\documentclass[11pt]{llncs}
\usepackage{graphicx}

\usepackage{times}
\usepackage{mathptm}
\usepackage{url}

\makeatletter
\def\hb@xt@{\hbox to }
\makeatother

\let\oldendproof\endproof
\def\endproof{\qed\oldendproof}

\begin{document}
\title{Single-Strip Triangulation of Manifolds with Arbitrary Topology}

\author{M.\ Gopi\thanks{gopi@ics.uci.edu} and David Eppstein\thanks{eppstein@ics.uci.edu}}
\institute{Department of Computer Science, University of California, Irvine}

\maketitle

\begin{abstract}
Triangle strips have been widely used for efficient rendering.
It is NP-complete to test whether a given triangulated
model can be represented as a single triangle strip, so many heuristics have
been proposed to partition models into few long strips. In this paper, we present a new algorithm
for creating a single triangle loop or strip from a triangulated model.
Our method applies a dual graph matching algorithm to partition the mesh into cycles,
and then merges pairs of cycles
by splitting adjacent triangles when necessary.
New vertices are introduced at midpoints of edges
and the new triangles thus formed are coplanar with their parent triangles,
hence the visual fidelity of the geometry is not changed.
We prove that the increase in the number of triangles due to this splitting is 50\%
in the worst case, however for all models we tested the increase was less
than 2\%.  We also prove tight bounds on the number of triangles needed for a single-strip representation of a model with holes on its boundary.  Our strips can be used not only
for efficient rendering, but also for other applications including the 
generation of space filling curves on a manifold of any arbitrary topology.
%\begin{classification} % according to http://www.acm.org/class/1998/
%\CCScat{I.3.5}{Computer Graphics}{Geometric algorithms, Triangulation, Stripification. } 
%\CCScat{G.2.2}{Graph algorithms}{Hamiltonian Path, Hamiltonian Cycle, Perfect Matching.}
%\end{classification}
\end{abstract}

\section{Introduction \label{sec:intro}}
Constructing strips from an input set of triangles has been an active field of research in
computer graphics and computational geometry, motivated
by the need for efficient rendering in the former and by traveling salesman and 
Hamiltonian path problems in the latter. Traditionally, triangle stripification 
research has been pursued along two extreme problem statements. 
At one end, the input model is considered unchangeable and 
algorithms are designed to test whether there is a Hamiltonian path in the
triangulation or to find as few strips as possible from the model.
At the other end, the input triangulation is completely ignored, and a new triangulation
is imposed on the input vertices in order to arrive at a single strip triangulation
even if it requires addition of new vertices. The work presented here
attempts to bridge the gap by finding a single strip triangulation from the input triangulation
by splitting the input triangles if necessary while guaranteeing that the
geometry of the input model is also retained.

\begin{figure}[t]
\centering
\includegraphics[height=3in]{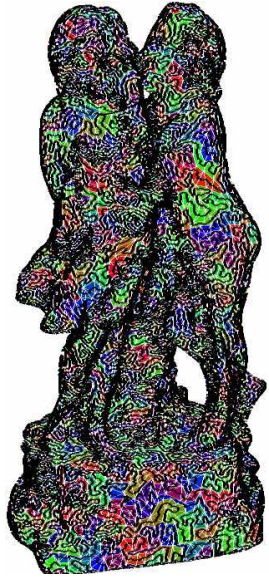}
\caption{Single cycle triangulation of the genus-two sculpture model with 50780 input and 51780 output triangles.}
\label{}
\end{figure}

Our motivation goes beyond the rendering requirement and
theoretical aspects of Hamiltonian paths. The advantages of having a single triangle
strip representation of a model enables a plethora of other geometric and topological
algorithms to be applied on the model. In this paper, we show one such application --
generating space filling curves on manifolds of arbitrary topology -- using the total linear 
ordering of triangles given by a Hamiltonian cycle.
 
 Triangle stripification algorithms can be categorized based on their input requirements. 
 The first category of algorithms takes only the vertices of the model as input and finds a 
 triangulation that would generate a single strip. The second category takes edges of a 
 polygon as input and triangulates the interior of the polygon, with or without the addition 
 of Steiner vertices, to create triangle strip(s) that covers the polygon.
 Typically, these two categories work
 only with data sets on a plane or a height field. 
 The third category takes triangles of the model as input and tries to build long
 triangle strips, not necessarily a single strip,  only using the input set of triangles. 
The third category works with 2D surfaces
 embedded in 3D.
The algorithm presented in this paper is a combination of all the above three categories and
 creates a single triangle strip from the model. In the triangle strip generated by our method, all 
input vertices are used as in the first category, Steiner vertices and hence more triangles 
 are added as in the second category, and finally, the geometry of the input triangulation is 
 retained as in the third category.

\subsection{Related Work}

% Dillencourt \cite{Dillencourt:92} showed that finding Hamiltonian cycles
% in Delaunay triangulation is NP-complete. 
% Since finding a single strip is equivalent to finding a Hamiltonian path, 
It follows from Steinitz' theorem and known results on NP-completeness of the
Hamiltonian cycle problem for cubic 3-connected planar graphs~\cite{GarJohTar-SJC-76}
that it is NP-hard to find a single strip, even for a model consisting of a triangulated convex polyhedron,
and known exponential-time algorithms for Hamiltonian cycles are not sufficient to find single strips for models of more than 100 triangles~\cite{Epp-WADS-03}.
Therefore, many algorithms simply attempt to find as few a strips as
possible from the input triangulation. SGI developed a program~\cite{AHB:90}
that produces generalized triangle strips using a heuristic that begins and ends strips on faces with few neighbors, so as to
reduce the number of isolated triangles. The classic STRIPE algorithm 
\cite{ESV:96a} makes a global analysis of the input triangle mesh, trying to find patches
that can be efficiently striped. Velho et al.~\cite{VFG:99} build and maintain triangle strips incrementally while creating a triangle mesh simultaneously. 
Chow \cite{Chow:97} builds strips by reusing the points added to previous strips
as often as possible.
Snoeyink and Speckmann \cite{SS:97} propose a stripification
algorithm specially designed for triangulated irregular network (TIN) models using
the spanning trees of the dual graphs of TIN models. Xiang et al.~\cite{XHM:99} 
decompose spanning trees of the dual graph into triangle strips. The tunneling method for
triangle strips in continuous level of detail meshes is proposed by Stewart~\cite{Stewart:01}.
Taking this a step further, Shafae and Pajarola~\cite{SP:03} propose dynamic triangle strip management for view-dependent mesh simplification and rendering algorithms. Demaine et al.~\cite{DemEppEri-WK-03} relaxed the definition of a triangle strip,
to allow adjacent triangles in the strip to share only a single vertex instead of an edge, and showed that any model consisting of triangles meeting edge-to-edge (possibly with boundary) admits such a relaxed strip.
Bogomjakov and Gotsman \cite{BG:02} investigate the ordering of triangles in order to reduce the number of vertex cache misses and develop methods to find triangle sequences that preserves the property of locality. Triangle strips are also important in geometric compression and transmission and is a by-product of these algorithms \cite{Rossignac:99,TG:98}. Here again, the input triangulation is usually not modified.

The hardness of finding
 Hamiltonian paths can be eased by minor variations of the problem statement.
 For example, algorithms presented in \cite{AHMS:96} avoid Delaunay triangulation of the
 planar point set and create a Hamiltonian path triangulation. Further, 
they  take as input a planar simple polygon, and check using its visibility graph
whether there exists a single strip triangulation of the polygon's interior. If not, 
such a triangulation is produced using Steiner vertices. They also prove that computing
a Hamiltonian triangulation for planar polygons with holes is NP-hard.
The QuadTIN method \cite{PAL:02} 
triangulates an irregular terrain point data set by adding Steiner vertices at quad-tree
corners to produce a dynamic view-dependent triangulation that can be 
traversed as a single strip. Given a quadrilateral mesh of a manifold, Taubin \cite{Taubin:02} splits each quadrilateral into triangles and orders them into a single strip. Unlike the above methods that take a quadrangulation or points on a plane or a
height field as input, our method uses the triangulation of manifolds 
of arbitrary topology. Further, even by adding new triangles, our algorithm does not 
change the input geometry (in terms of visual fidelity), whereas the above methods prescribe a completely new triangulation that includes the input point set.

\subsection{New Results}
\begin{figure*}[t]
\includegraphics[width=0.24\columnwidth]{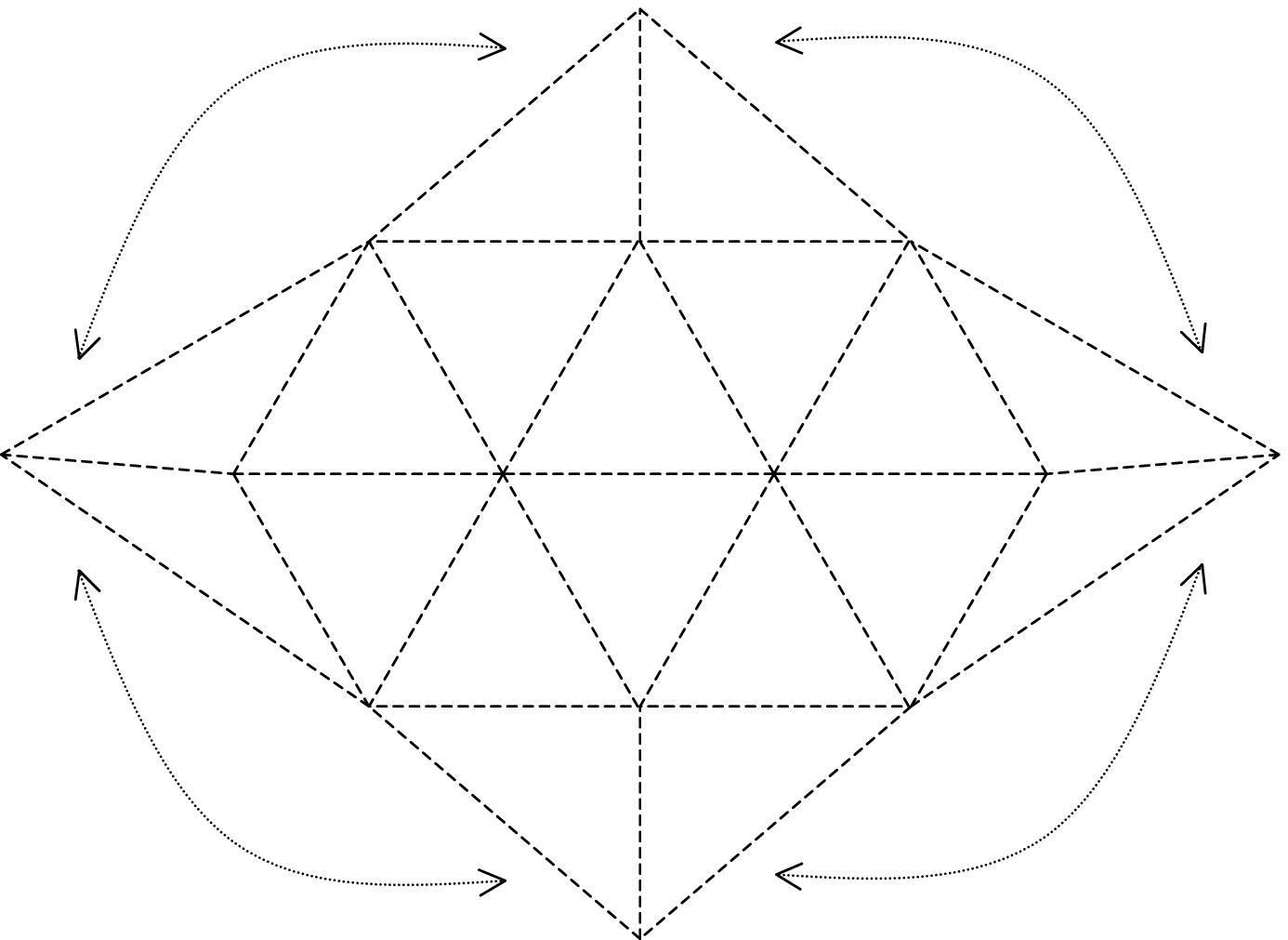}
\includegraphics[width=0.24\columnwidth]{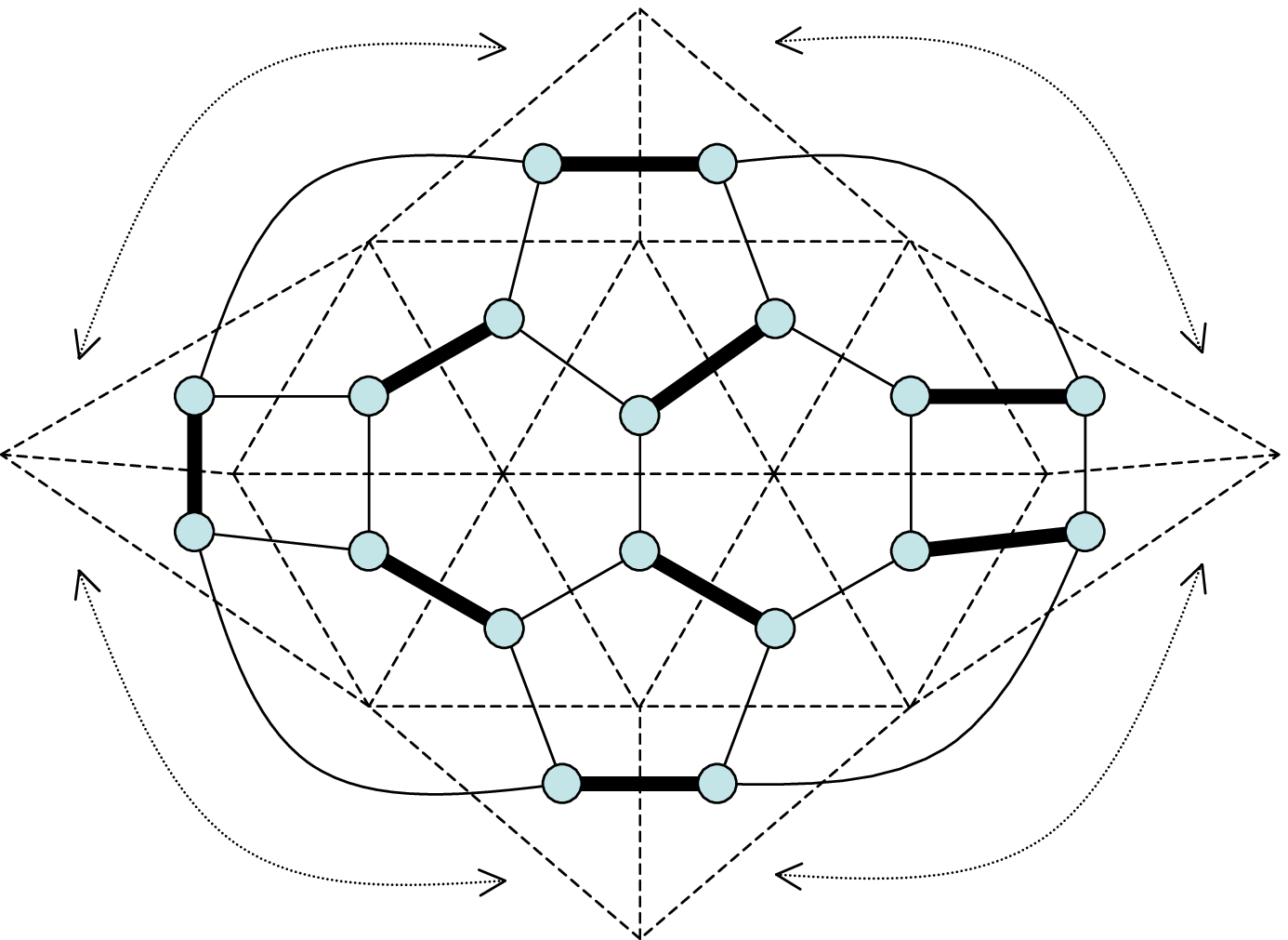}
\includegraphics[width=0.24\columnwidth]{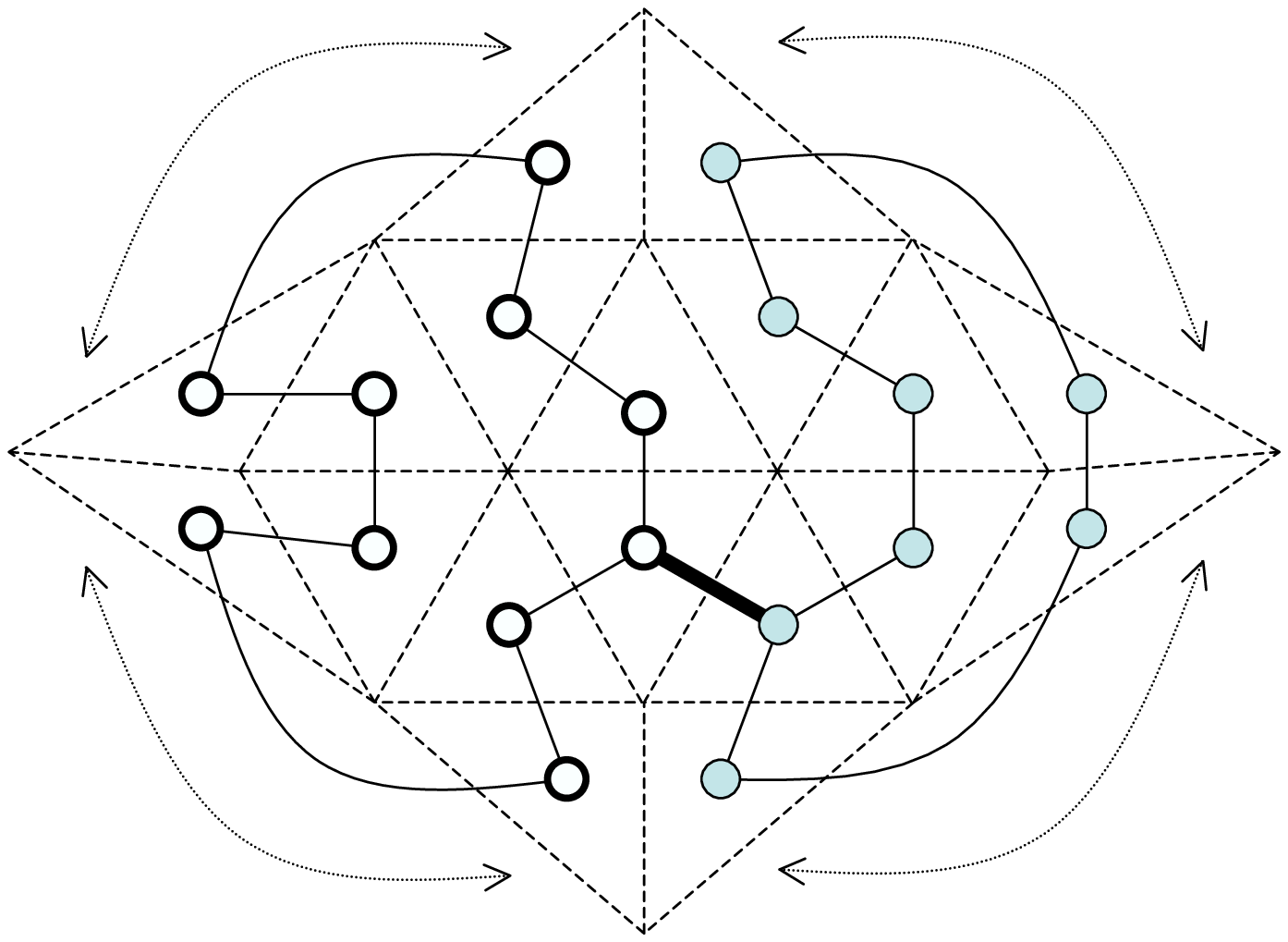}
\includegraphics[width=0.24\columnwidth]{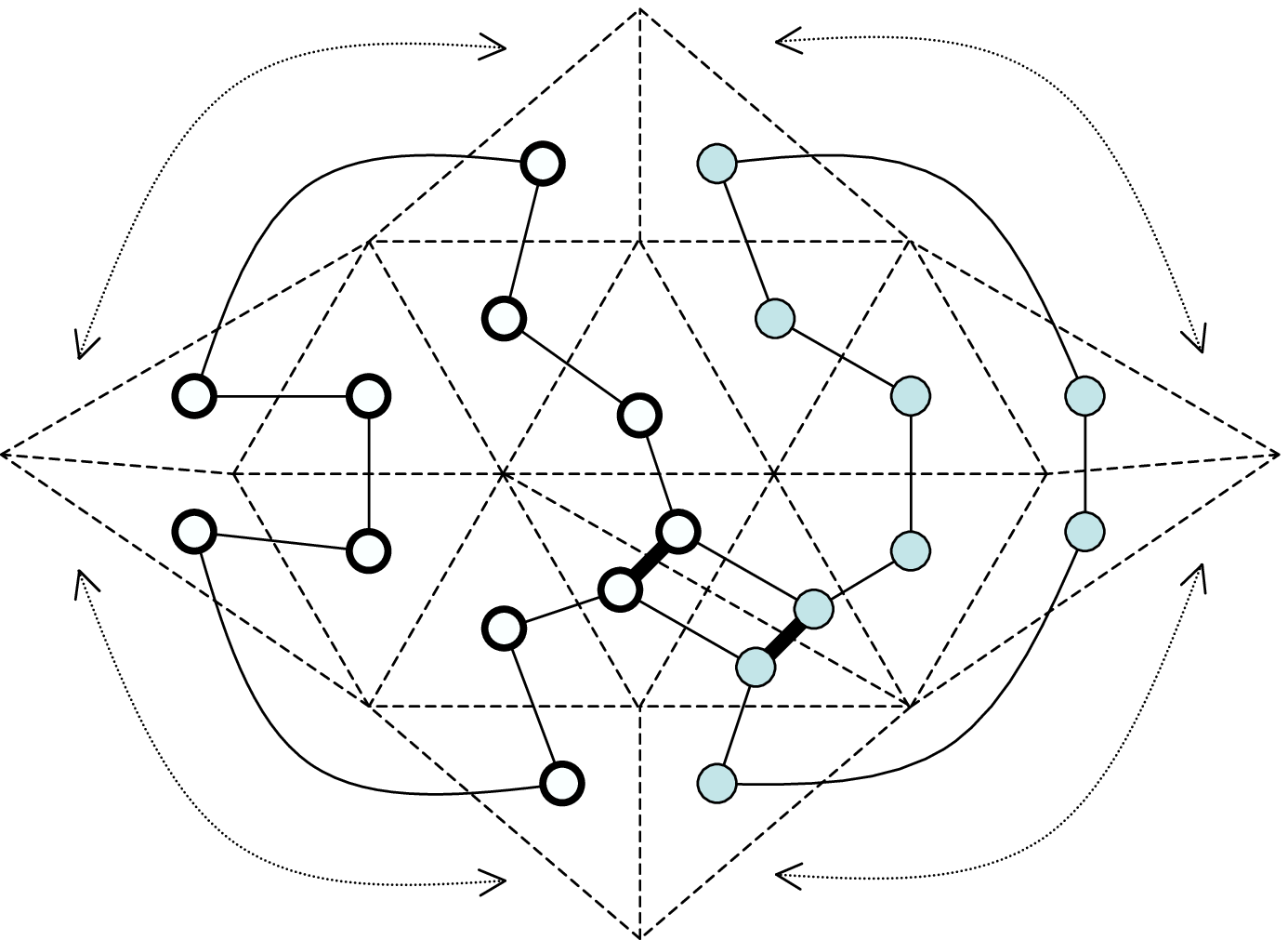}
\caption{(a) The given triangulation. Assume that the boundary edges are adjacent to
each other as shown by dotted arrows and all the boundary triangles share a common vertex. The mesh shown is a manifold with genus 0. (b) The dual degree three graph and a perfect matching
shown by dark edges. (c) If we remove all the matched edges we get disjoint cycles consisting of
only unmatched edges. Two such cycles, one with unfilled and the other with filled graph nodes are
shown. The union of these cycles cover all the given vertices in the dual graph and hence the triangles in the given mesh. These disjoint
cycles are connected to each other by matched edges. We construct a spanning tree of these disjoint cycles
and hence choose matched edges that connect these cycles. (d) The triangle pair corresponding to chosen matched edges in the tree are split creating two new triangles. Matching is toggled around
the new vertices  
resulting in a triangulation with a Hamiltonian cycle of unmatched edges as shown in the figure.}
\label{fig:matching}
\end{figure*}
\begin{figure*}[t]
\includegraphics[width=0.33\columnwidth]{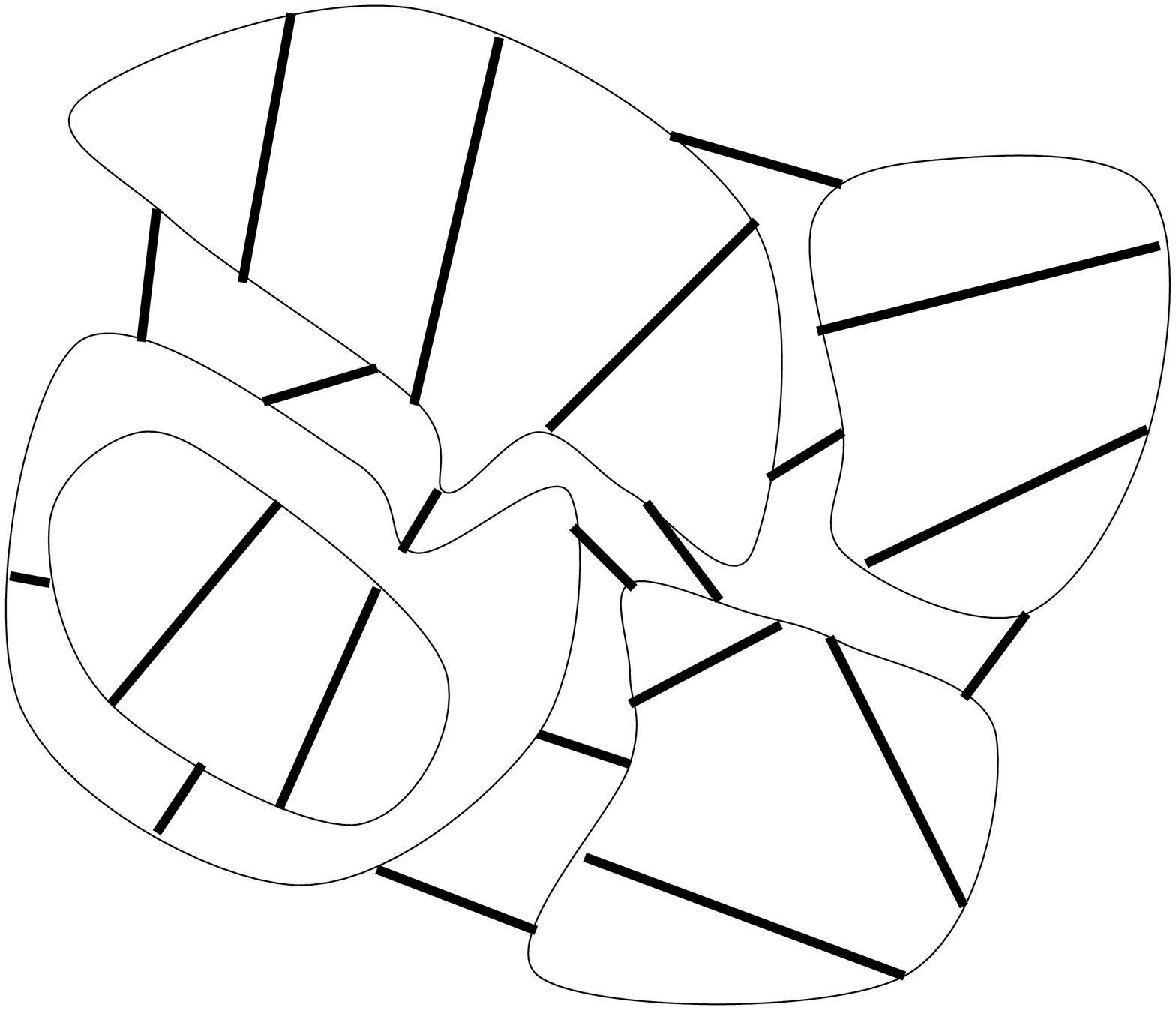}
\includegraphics[width=0.33\columnwidth]{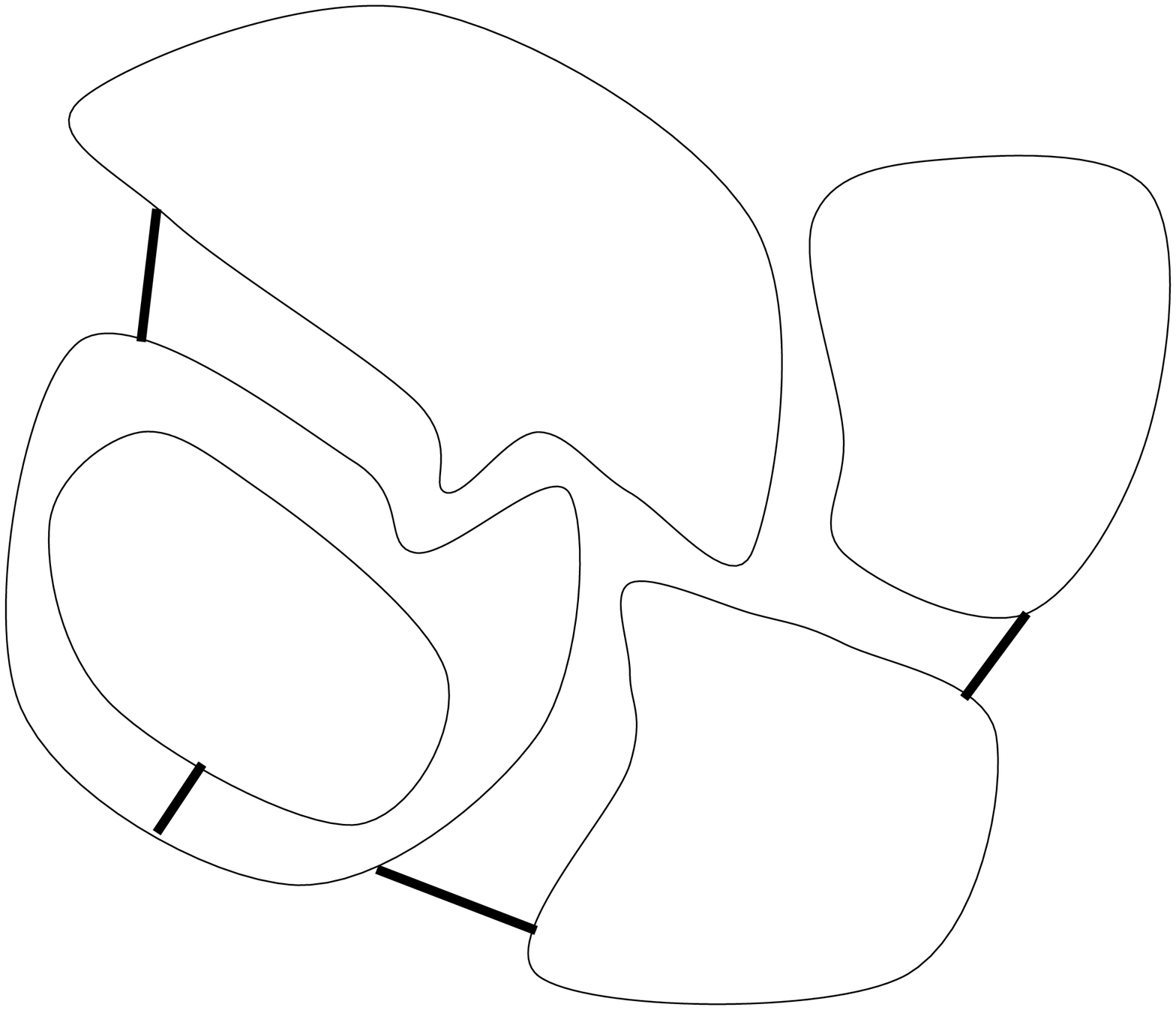}
\includegraphics[width=0.33\columnwidth]{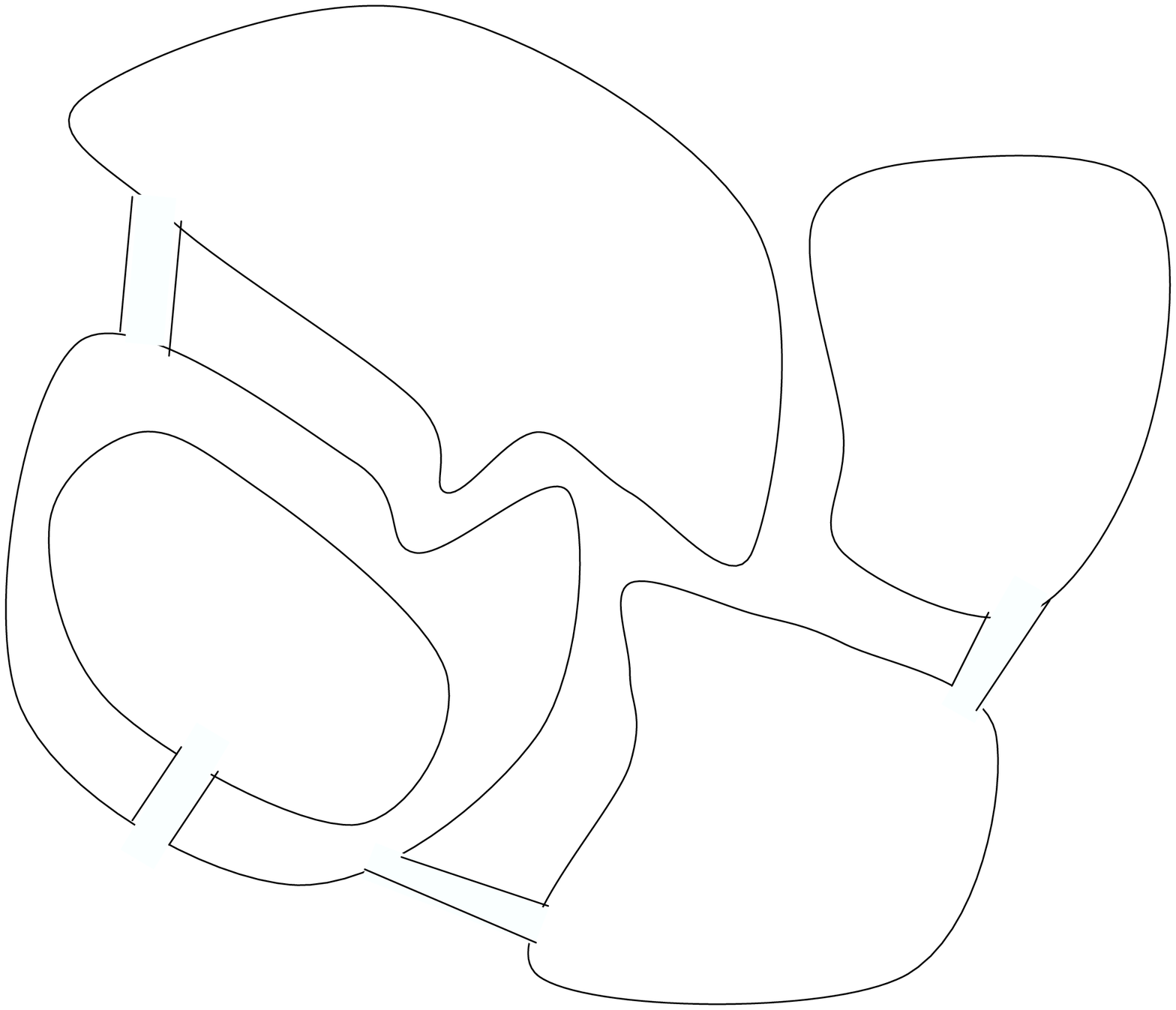}
\caption{(a) An example of a graph of disjoint cycles of unmatched edges. These cycles are connected to each other by matched edges. (b) A spanning tree of cycles is constructed 
using selected matched edges. (c) These matched edges are split to create a Hamiltonian cycle.}
\label{fig:cycletree}
\end{figure*}

Our main result is a new method for subdividing a triangulated model and finding a single triangle strip in the subdivided model.  We show theoretically that this method is guaranteed not to increase the number of triangles in the model by more than 50\%, but in our experiments the increase was at most 2\%.  In order to estimate  how tight our worst case bounds are, we also construct a lower bound, consisting of an infinite family of triangulated models in which any method for subdividing the model to produce a single triangle strip must increase the number of triangles by at least 5.4\%.

We also consider the problem of producing a single strip by subdividing a model having a triangulated boundary with holes.  For such a model, with $n$ triangles, we show a tight bound of $3n-4\log_2 n\pm O(1)$ on the number of triangles in the resulting strip.  Therefore, starting with a watertight model is of considerable benefit in stripification.

\section{Hamiltonian Cycle Stripification \label{sec:algo}}
In this section we describe our algorithm to create a Hamiltonian cycle from
the given triangulation.
The fundamental technique we use to arrive at a Hamiltonian cycle is 
a perfect matching algorithm. 
A {\em matching} in a graph $G=(V,E)$ is a 
subset M of the edges E such that no two edges in M share a common 
end node. A {\em perfect matching} $M$ in $G$ is a matching such that each node 
of $G$ is incident to an edge in $M$.  (See Figure \ref{fig:matching}(b).)
For a bridgeless graph in which every vertex has 
degree three, there always exists a perfect matching~\cite{P:1891}.
Such a matching can be found in time
$O(n)$ for planar graphs~\cite{BieBosDem-Algs-01}
or $O(n\log^4 n)$ in general,
where $n$ is the number of input vertices;
the latter bound can be further improved to
$O(n\log^3 n\log\log n)$ using recent results of Thorup~\cite{Tho-STOC-00}.
In our case, we are interested in the dual graphs
of triangulated manifolds; such graphs are bridgeless (in fact, 3-connected) and
have degree three.  A perfect matching in this dual graph will pair every triangle with exactly
one of its adjacent triangles.

Given a perfect matching for a degree three graph, there are two {\em unmatched 
edges} incident on every graph vertex. The set of all unmatched edges 
forms a collection of disjoint cycles, the union of which covers the complete vertex set of
the graph (Figure \ref{fig:matching}). These disjoint cycles are adjacent to each 
other across matched edges. Let us construct a graph called the {\em cycle graph} in which the nodes correspond to the disjoint cycles of this collection, and two nodes share an edge
whenever the corresponding two cycles are adjacent to each other across a matched edge.
From the cycle-graph, we construct a spanning tree of cycles (Figure \ref{fig:cycletree}). 
Considering this graph as the dual of our triangulation,
the disjoint cycles are triangle strips (loops) and the matched edges in the tree are
adjacent triangle pairs. As shown in Figures \ref{fig:matching}(d) and \ref{fig:cycletree}(c), 
we split each of these 
matched triangle pairs corresponding to the matched edges in the tree, to form
a single cycle connecting all the triangles in the manifold. 

If $k$ is the number of disjoint  cycles then we need $(k-1)$ matched edges 
to form a spanning tree (Figure \ref{fig:cycletree}).
Triangle pairs corresponding to these $(k-1)$ matched edges have to be split introducing
$2(k-1)$ new triangles.
Since the number of triangles in each cycle cannot be less than three, $k\leq \frac{n}{3}$.
This worst case scenario results in $n+2\frac{n}{3} = 1.66n$ triangles in the Hamiltonian cycle.

\subsection{Eliminating Three-Cycles}

\begin{figure}[t]
\centering
\includegraphics[width=0.75\columnwidth]{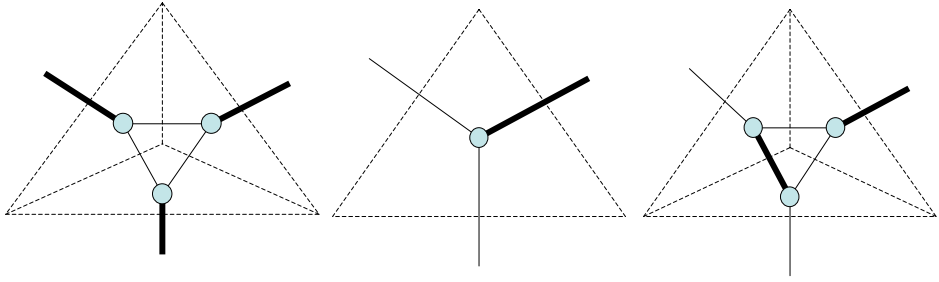}
\caption{(a) A configuration of three triangles surrounding a degree-three vertex. (b) From the input mesh, all degree-three vertices
are removed before finding the matching. The typical result after the matching is shown here. (c)
The removed triangles are reinserted with appropriate matching between them. This local change does not change the matching outside the configuration. This process results in a perfect matching such that the unmatched edges have no three-cycles.}
\label{fig:three-cycle}
\end{figure}

As discussed above, the worst case for our algorithm arises when the cycles of unmatched edges have length exactly three.
We eliminate the possibility of occurrence of such three-cycles, improving our worst-case guarantees on the number of subdivisions, using the following observation. 
Cycles consisting of three triangles are formed only if there exists a configuration of three mutually adjacent triangles surrounding a degree-three vertex as shown in Figure \ref{fig:three-cycle}(a).  We temporarily simplify the mesh by repeatedly removing the central vertex from each such configuration, replacing the configuration by a single triangle (Figure \ref{fig:three-cycle}(b)); in the dual graph of the mesh, this corresponds to a $\Delta$-$Y$ transformation in which a three-cycle is contracted to a single point.
It is not a concern that this simplification may lead to self-intersecting geometry in the mesh.
We can test whether any triangle belongs to a three-cycle by examining a constant number of nearby triangles, so the total time for this transformation is linear.

Once all three-cycles are removed, we then apply Petersen's theorem to find a perfect matching in the simplified mesh.  A typical result after such a 
matching is shown in Figure \ref{fig:three-cycle}(b). We then add the removed three-triangle configurations back one at a time into this `matched' mesh as shown in Figure \ref{fig:three-cycle}(c). 
At each step, two of the triangles in the configuration are matched to each other, so that globally we retain a perfect matching with no three-cycles among the unmatched edges.

By means of this optimization, we are able to prove the following result.

\begin{theorem}
For any triangulated model with $n$ triangles,
we can find in time $O(n\log^3 n\log\log n)$ a subdivision of the model,
and a single triangle strip for the subdivision,
in which the subdivision has fewer than $3n/2$ triangles.
If the model has the topology of the sphere then
the time for finding a single strip subdivision
can be further improved to $O(n)$.
\end{theorem}

\begin{proof}
As discussed above, we find a perfect matching in the dual of the input model,
such that the unmatched edges form cycles with no three-cycle.
Therefore, there can be at most $n/4$ cycles, $n/4-1$ edges selected in
the spanning tree of the cycle graph, and $2(n/4-1)$ subdivisions.
The total number of triangles is thus at most $3n/2-2$.
The processes of removing and restoring three-triangle configurations,
and of finding a spanning tree for the cycle graph and using it to select a set of subdivisions
to perform, all take only linear time, so the total time is bounded by the algorithm for
finding perfect matchings.
\end{proof}

Although this worst case upper bound on additional 
number of triangles is 50\% of the input number of
triangles, in practice additional triangles is less than 2\%. We achieve this result after using the 
further optimization described below. The results on lower bounds on additional triangles for
both manifolds and manifolds with boundaries are given in the Appendix.

\begin{figure}[t]
\centering
\includegraphics[width=0.6\columnwidth]{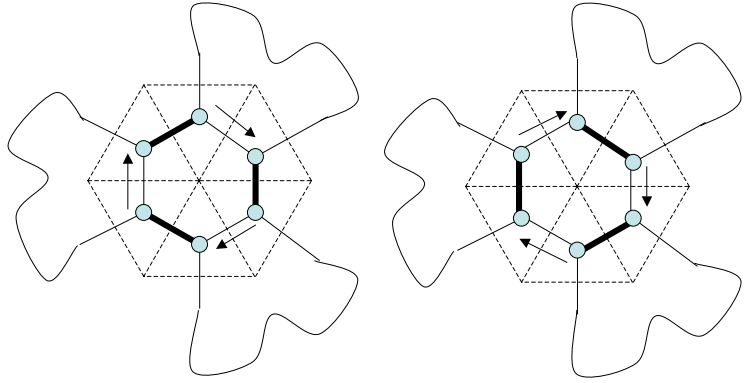}
\caption{(a) A nodal vertex with (six) even number of incident triangles and triangles belonging to
three unique cycles. (b) By switching the matched and unmatched edges, all these cycles can be
merged to a single cycle.}
\label{fig:nodal-vertex}
\end{figure}

\begin{figure}[t]
\centering
\includegraphics[width=0.66\columnwidth]{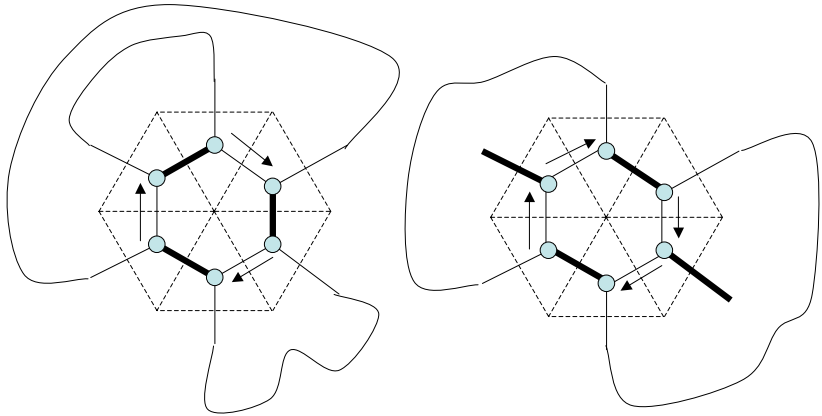}
\caption{Examples of non-nodal vertices. In both the examples, there are six incident triangles but only two unique cycles.}
\label{fig:non-nodal-vertex}
\end{figure}

\subsection{Merging Cycles Around Nodal Vertices}

The goal of this optimization is to increase the length of the disjoint cycles by merging many cycles
without any triangle splits.  Assume that we have already constructed a perfect matching, and partitioned the triangles of the input mesh into disjoint cycles.  We classify a mesh vertex $v$ as a {\em nodal vertex} if it satisfies the 
following conditions: $v_n$, the number of triangles incident on $v$ is even 
and the total number of unique disjoint cycles that these incident triangles belong to is $\frac{v_n}{2}$. 
An example of a nodal vertex with six incident triangles and three unique incident cycles is shown in
Figure \ref{fig:nodal-vertex}.  The neighborhood of every nodal vertex is modified such that the
matched and unmatched triangle pairs are toggled. This merges all the incident cycles into one cycle.
Examples of non-nodal vertices are shown in Figure \ref{fig:non-nodal-vertex}.

If we use a union-find data structure to keep track of which triangles belong to which cycles,
we can test whether any mesh vertex is nodal using a number of union-find queries proportional to the degree of the vertex, so the total time for the optimization is $O(n\alpha(n))$ where $\alpha$ is the extremely slowly growing inverse Ackermann function.

Once this optimization is performed, we form the cycle graph of the remaining cycles, construct a spanning tree, and use the tree to guide triangle subdivisions as before.
This optimization step typically significantly reduces the number of subdivisions that must be performed, but we have no theoretical guarantees on its performance. The results of these optimizations on 
various models are shown in Table \ref{table:results}.

\begin{table}[h]
\begin{center}
\begin{tabular}{|l|r|r|r|} \hline
Model & Input Tris & Output Tris & \% Increase \\ \hline

Torus&400&406&1.5\\
Sphere&480&484&0.8\\
Goblet&1000&1016&1.6\\
Eight&1536&1554&1.1\\
Sculpture&50780&51780&1.9\\
Fandisk&12946&13134&1.4\\
Horse&96966 &98552&1.6\\
Shoe&156474&158980&1.6\\ \hline
\end{tabular}
\end{center}
\caption{Results of our algorithm on different models. Note that the increase in number of
triangles is consistently less than 2\% of the input number of triangles.}
\label{table:results}
\end{table}%

\section{Matching Implementation Details } 
\label{sec:implement}

We describe the implementation details only for the perfect matching phase of our algorithm, since the other parts of the algorithm are straightforward.

Rather than using the theoretically efficient but somewhat complex
$O(n\log^3 n\log\log n)$ algorithm to construct perfect matchings
in the dual graphs of our models,
we used a general purpose graph maximum matching algorithm,
Edmonds' blossom-contraction algorithm~\cite{Edm-CJM-65}.
This algorithm repeatedly increases the size of a matching by one edge,
each time performing a breadth-first search using a union-find data structure to keep track
of certain contracted subgraphs called {\em blossoms}.
Therefore, if the number of repetitions is $k$, the total time is $O(kn\alpha(n))$.

In our application of this matching algorithm, the number of edges in the resulting
matchings is always exactly $n/2$, so if we applied Edmonds' algorithm starting from an empty matching
we would take more than quadratic time, imposing strong limits on the size of the models
we could handle.  To reduce this time penalty, we precede Edmonds' algorithm by a greedy
matching phase, using the {\em degree-one reductions} and {\em degree-two reductions} described by Karp and Sipser~\cite{KarSip-FOCS-81} and Magun~\cite{Mag-JEA-98}.
This greedy matching phase typically matches 99.9\% of the triangles of our input models,
significantly reducing the time requirements for our perfect matching algorithm.

Our implementation of the matching portion of our strip finding algorithm
is written in Python, a relatively slow interpreted language.
On an 800 MHz Apple PowerBook, our code 
took approximately three minutes to find
a perfect matching for the
90K-triangle horse model
and took approximately seven minutes
to find a perfect matching for the 150K shoe model.
We expect that significant additional speedups could be obtained by rewriting our code in a faster compiled language, and by incorporating the more sophisticated greedy matching heuristics described by Magun~\cite{Mag-JEA-98}.

\section{Applications}
\label{sec:application}

\begin{figure}[t]
\centering
\includegraphics[width=0.4\columnwidth]{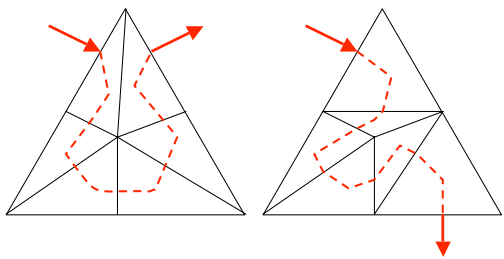}
\caption{Depending on the direction of the cycle, two ways of subdividing a triangle 
are shown. Triangle vertices are assumed to be ordered in counter-clockwise direction.
The space curve will pass through the edge centers and centroids of the triangles as shown. 
Recursive subdivisions would yield denser curves and eventually would fill the space of the 
given triangle.}
\label{fig:subdivision}
\end{figure}
Hamiltonian strip triangulation has many applications in rendering.
One such application we elaborate here is a procedural method to generate 
space filling curves \cite{Sagan:94} on the manifolds. Space filling curves are ideal for hierarchical
indexing of a higher dimensional space with a single parameter curve. 
They have tremendous applications in 
many fields and are used to solve various problems on 2D images including  
 contact searching \cite{DHLTW:00}, parallelizing finite 
element grid generation \cite{BZ:00},  out-of-core visualization algorithms for
massive mesh simplification \cite{LP:02} and volume rendering\cite{PF:01}, 
mesh indexing \cite{NS:96}, image compression \cite{PW:00},
dithering \cite{BV:95}, half-toning \cite{ZW:93}, etc.
Interesting applications of space filling curves in designing geometric data structures
are detailed in \cite{ARRWW:97}. Popular space filling curves include Hilbert curves 
and  Lebesgue's curves. A method proposed by Bartholdi and Goldsman \cite{BG:01} proposes a space filling curve on a vertex connected triangulation. Here we introduce a space filling curve that fills any (edge connected) triangulated manifold of any arbitrary topology. 
We believe that this procedural method to fill the surface with a single space-filling curve would enable researchers to do many of the geometric operations listed above, directly on the surface of 3D objects.

\begin{figure*}[t]
\includegraphics[width=0.24\columnwidth]{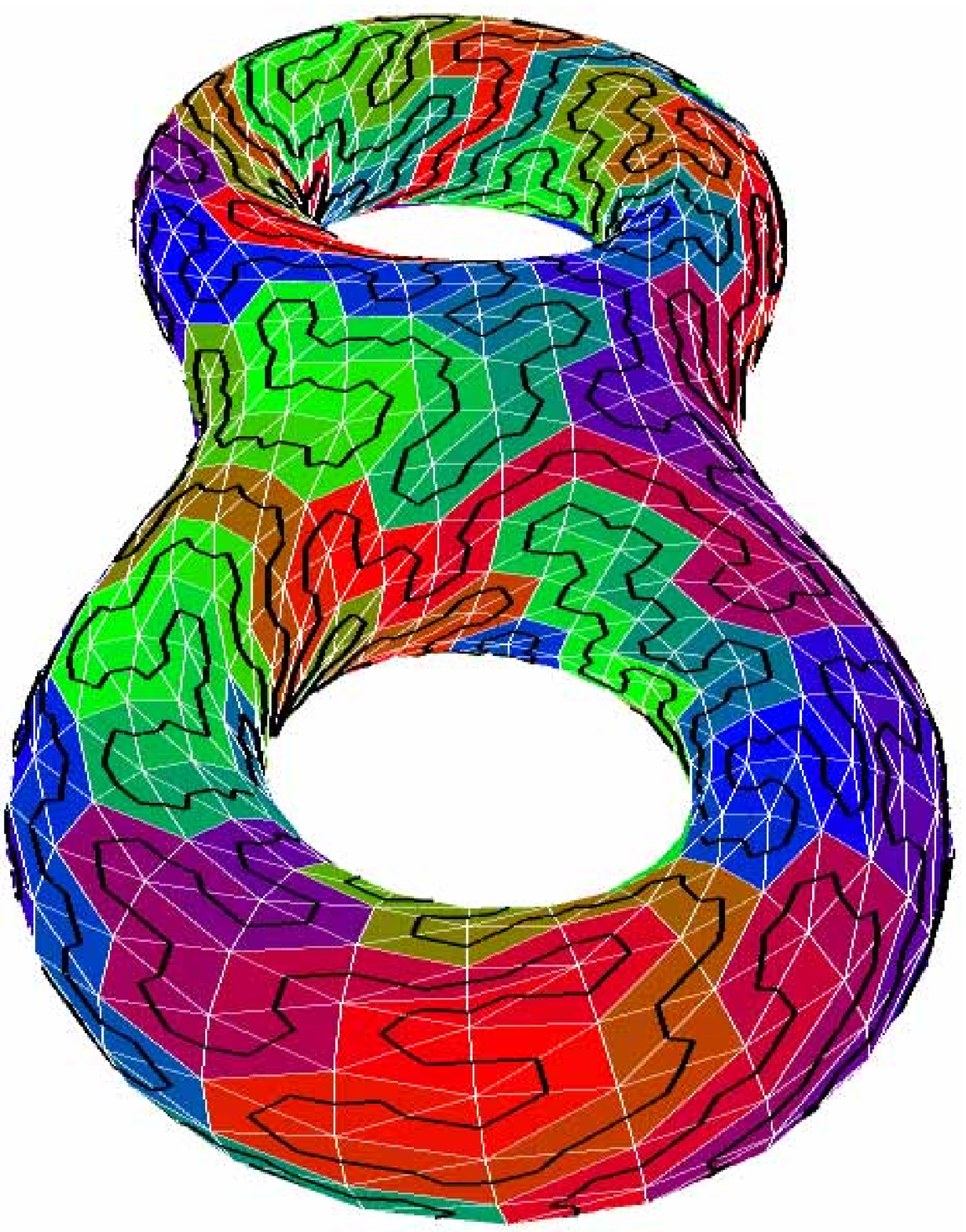}
\includegraphics[width=0.24\columnwidth]{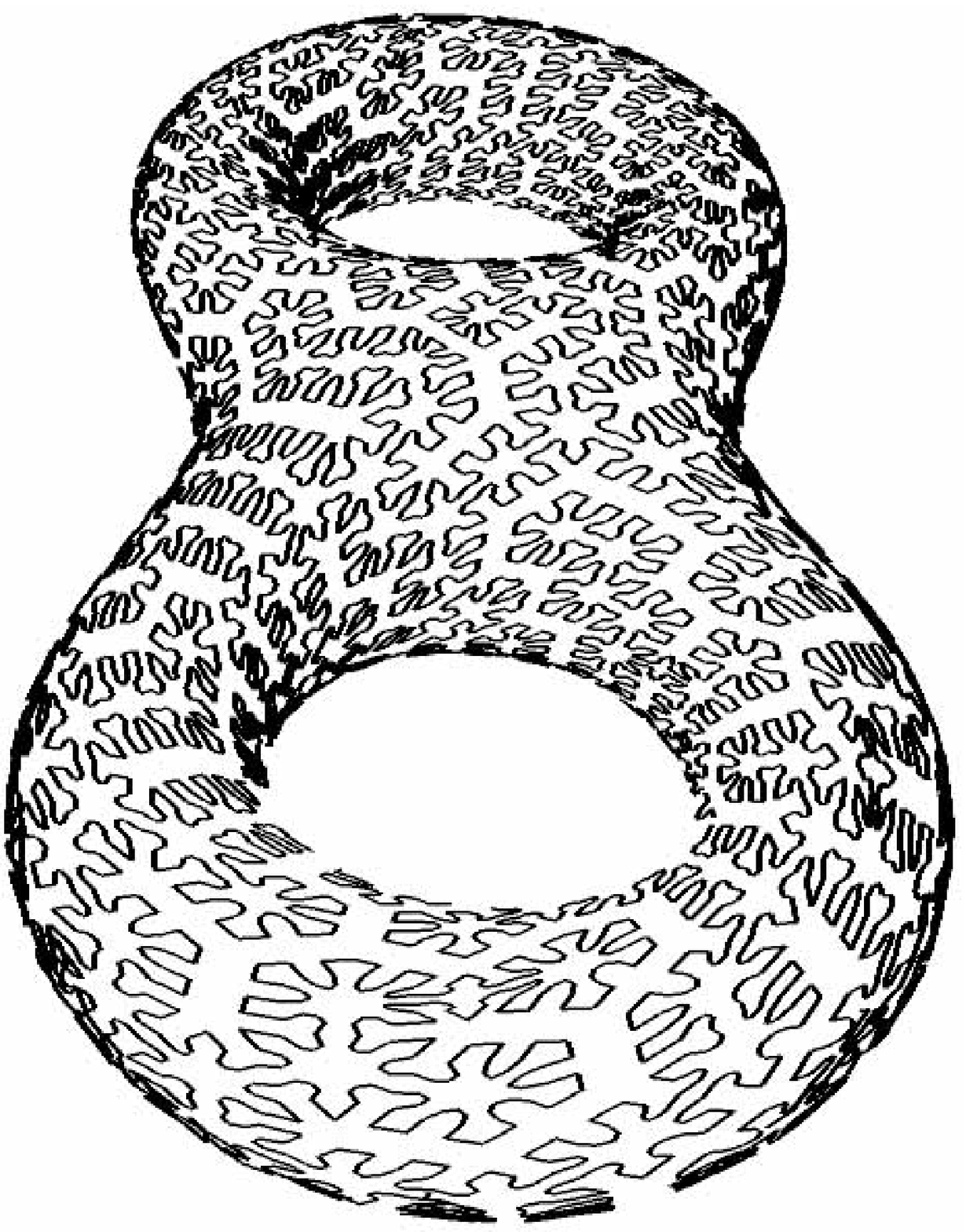}
\includegraphics[width=0.24\columnwidth]{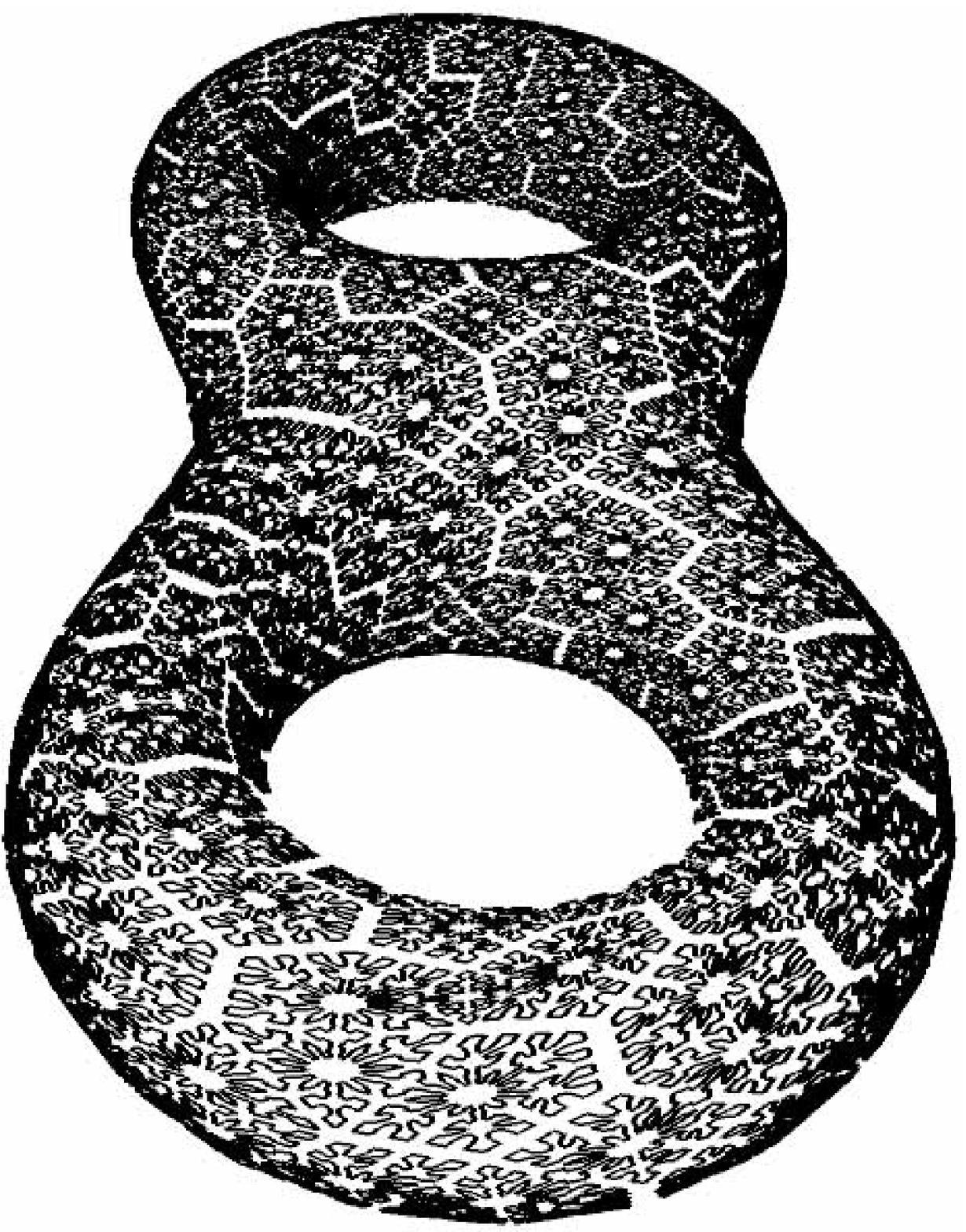}
\includegraphics[width=0.24\columnwidth]{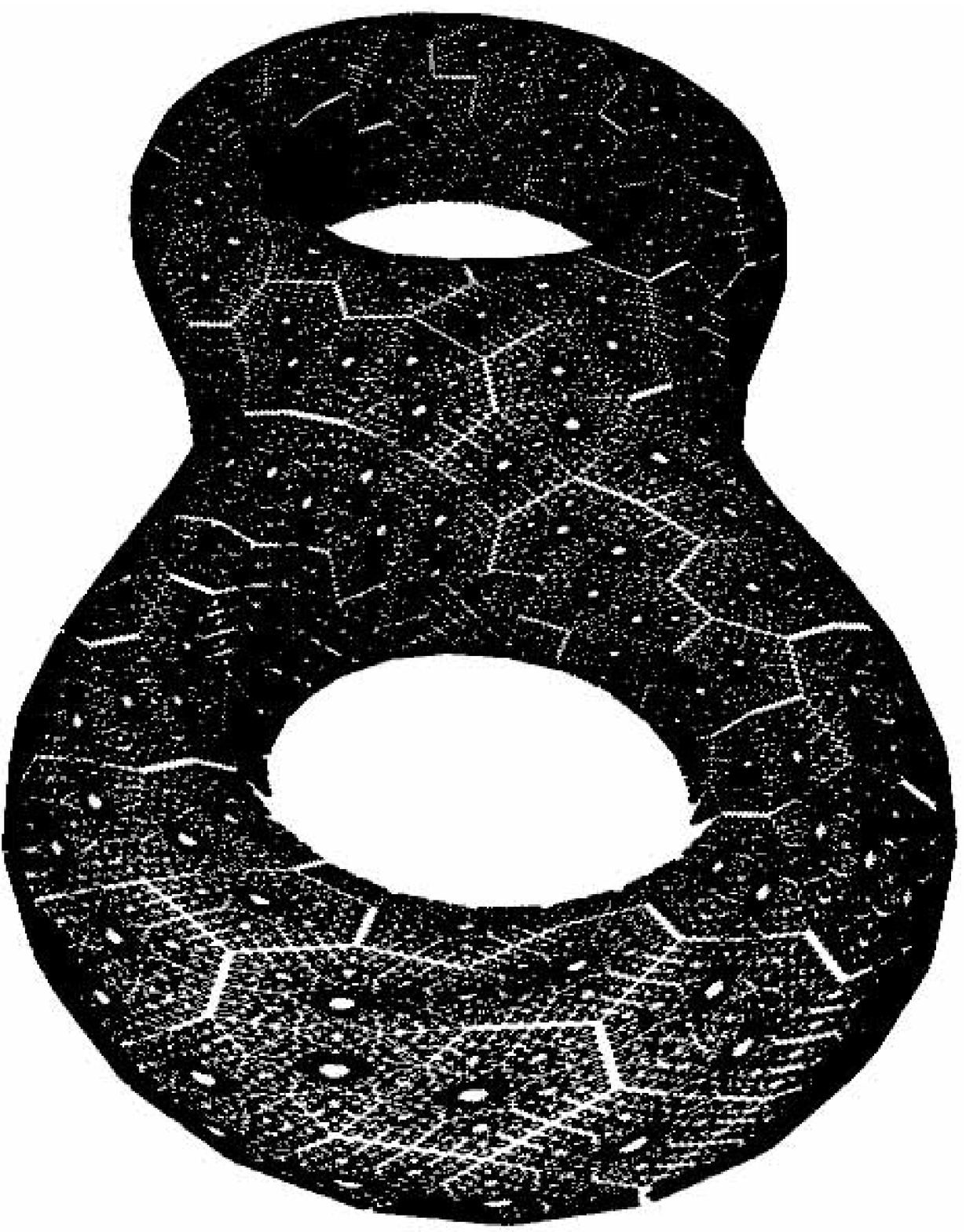}
\caption{A double torus with multiple iterations of space filling curves.}
\label{fig:space-filling}
\end{figure*}

Since the cycle produced by our triangle strip method covers the whole model, a curve passing 
through the triangles of the strip, in order, would cover the whole model. If this curve fills each
triangle on its way, it fills the space of any 2-manifold.
Like any other space filling curve, we define our curve by recursive subdivision of the triangle. 
The subdivision should ensure that the new vertices converge to the vertices and centroid of the
triangle. Hence edges have to be split by the subdivision. 
To ensure consistent subdivision of all the triangles and continuity of the curve across triangle
boundaries, we impose a direction to the Hamiltonian
cycle and also orient the manifold. Based on this direction of the cycle and the orientation of 
a triangle, 
we define two cases of subdivisions of the triangle, as shown in Figure \ref{fig:subdivision}.
For any given point on the manifold's surface,
each level of subdivision reduces the diameter of the triangle containing that point by a constant
factor, so the subdivision process produces in the limit a curve that passes arbitrarily close to the point
and is therefore truly space-filling curve.
Results of subdivision and space filling curves over multiple iterations are shown in Figures \ref{fig:space-filling}.
As far as we know, this is the first procedural method to produce space filling curves on manifolds of arbitrary
topology.

\section{Conclusion}
\label{sec:conclusion}
We have presented a new algorithm to strip the given triangulation of a manifold
into a single cycle covering all the input triangles. In the  process we split 
a small number of pairs of adjacent triangles at the midpoint of their common edges. 
This addition of triangles will maintain the geometric visual fidelity of the original model.
We have proven theoretical bounds on the output size of this algorithm, and also
shown experimentally that in practice the algorithm performs much better than the theoretical bounds would suggest.
%; detailed experimental results are presented below in Table~\ref{table:results}.
We also presented one of the many applications of Hamiltonian cycle triangulation, namely,
generating space filling curves on 2-manifolds with arbitrary topology. 

There are many future directions to this project. In this current project, we
have not considered the swap operations used in triangle strip rendering. New perfect matching
algorithms that would take these constraints into account have to be developed. Total linear
ordering of triangles is another powerful tool, but the Hamiltonian cycle provided by our algorithm
is more than a total linear ordering, so perhaps some reduction in output size can be achieved
by finding a path instead of a cycle.
%We have yet to explore other applications that use this linear ordering of triangles on manifolds.
We believe that the existence of,  and a procedural method to generate, a Hamiltonian cycle
triangulation will spark varied interests in the research community.

\begin{figure*}[t]
\includegraphics[height=2in]{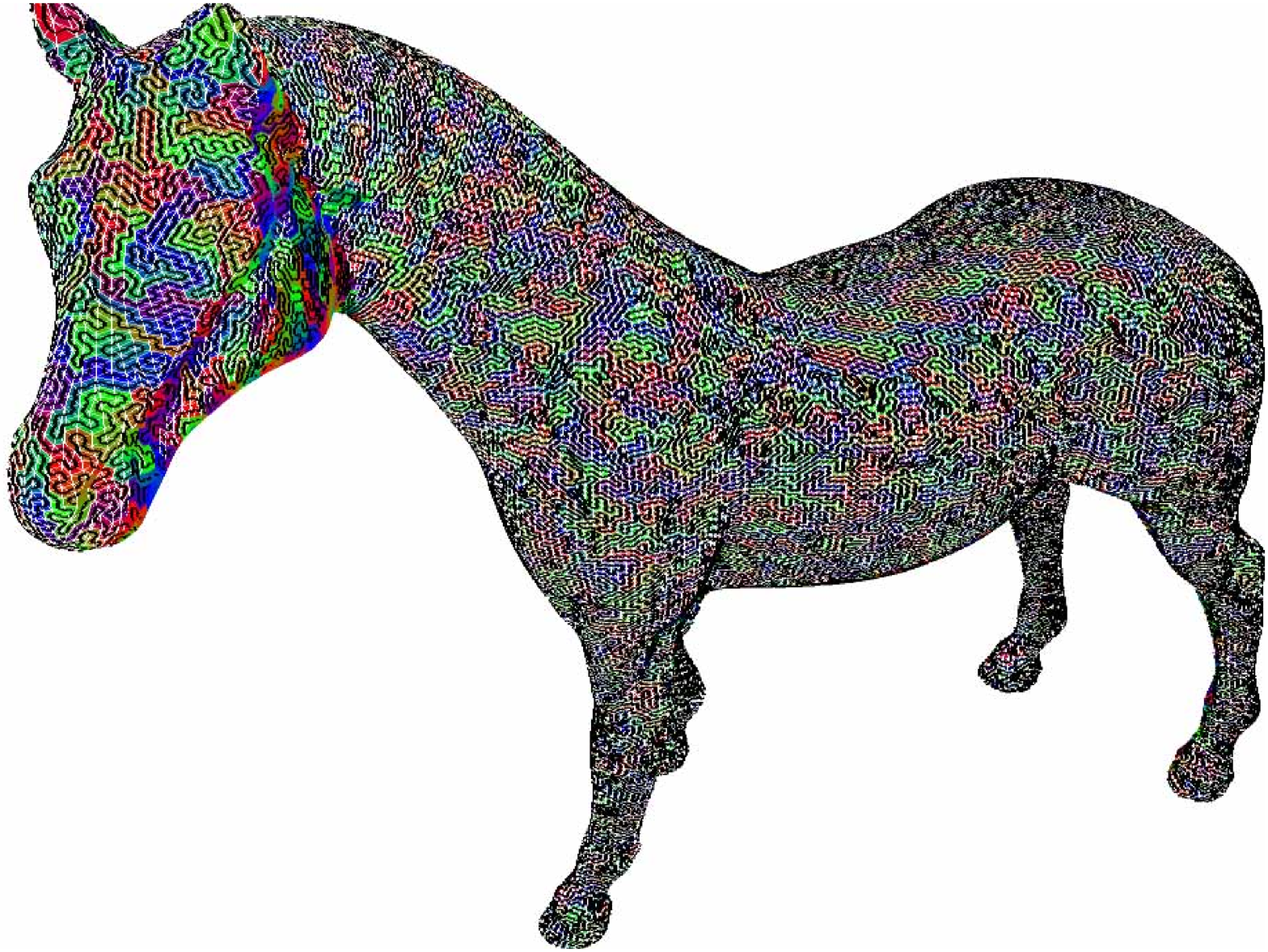}
\includegraphics[height=2in]{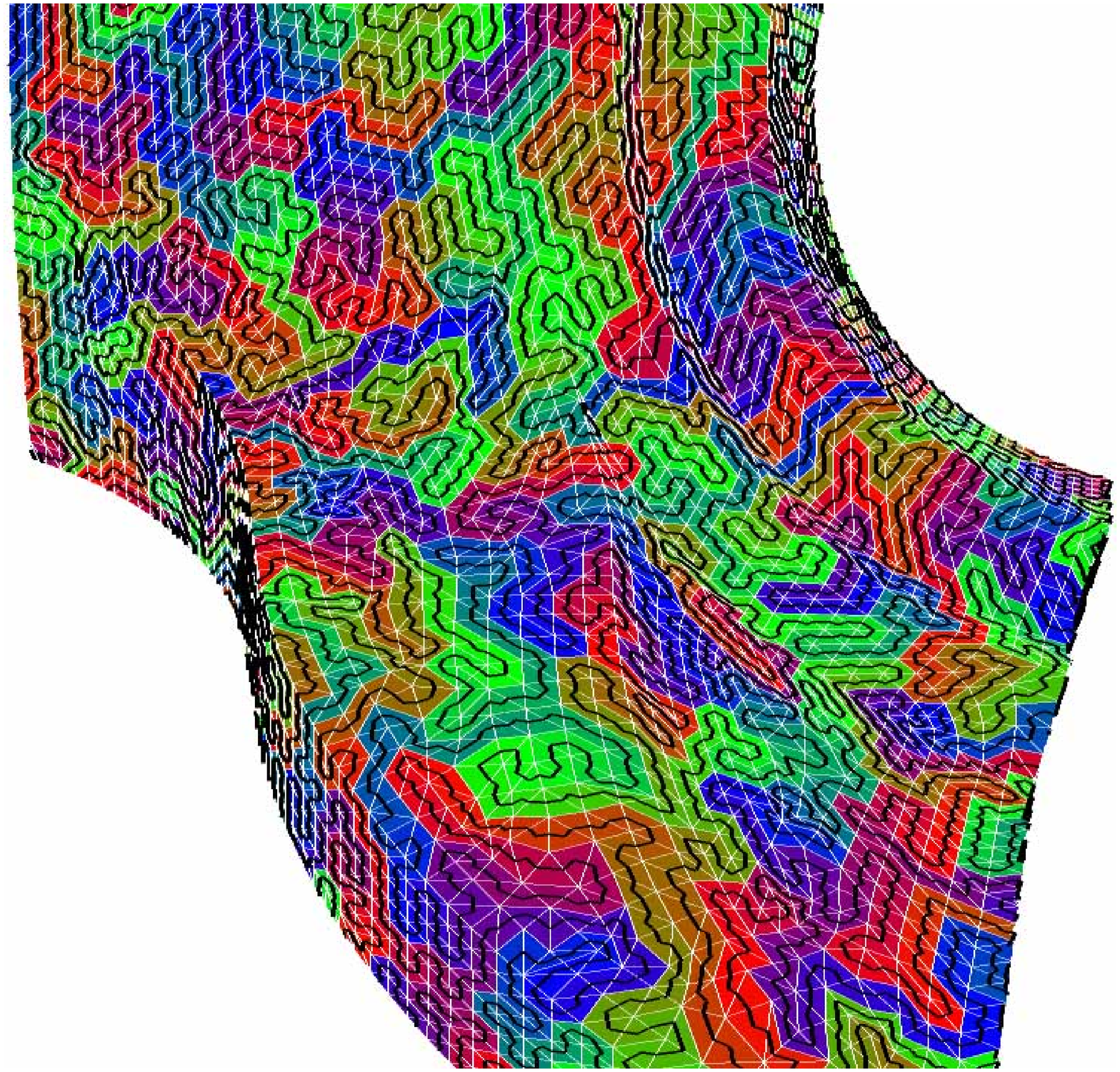}
\includegraphics[height=2in]{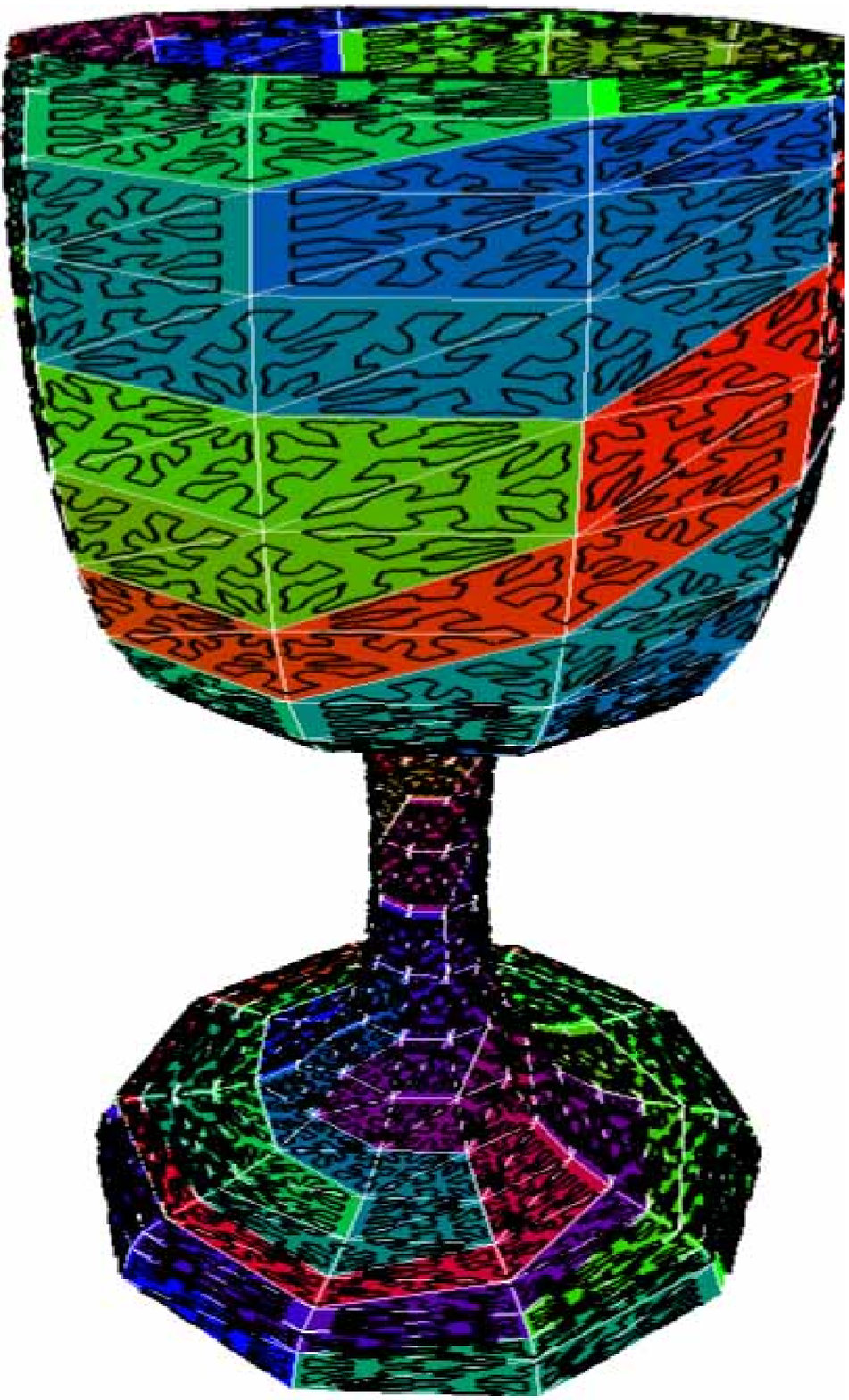}
\caption{Horse model with 96966 input and 98552 output triangles. Fandisk model with 12946 input
and 13134 triangles. Goblet with 1000 input and 1016 output triangles shown here with 
two iterations of space filling curve.}
\label{fig:results}
\end{figure*}

\bibliographystyle{eg-alpha}
\bibliography{biblio,morestriprefs}

\section*{APPENDIX: Lower Bound Analysis}

Although we proved a worst case bound of 50\% on the increase in the number of triangles of a model due to our subdivision process, our experimental results show that increases of only 2\% are more typical, and indicate to us that it may be possible to significantly improve our worst case bound.
How much improvement is possible?  To test this, we provide here a lower bound, showing the existence of models in which no subdivision method for producing single triangle strips can be guaranteed to achieve better than a 5.4\% increase in the number of triangles.

\subsection*{Lower Bounds for Manifolds}
The starting point of our lower bound is a result of Holton and McKay~\cite{HolMcK-JCTB-88},
that there exist non-Hamiltonian 3-connected 3-regular planar graphs with 38 vertices.
The dual of such a graph is a mesh of 38 triangles that can be realized as a convex polyhedron
and that has no cyclic strip of triangles.  By connecting many such meshes together, we form a more general lower bound.

\begin{theorem}
There exists an infinite family of triangulated convex polyhedra, such that any single triangle strip formed by subdividing an $n$-triangle polyhedron in the family must  have at least $39n/37-4$ triangles in the strip.
\end{theorem}

\begin{proof}
Let $H$ be the planar dual of the Holton-McKay graph; $H$ is a planar 3-connected graph with 38 triangular faces.
For any $k$, let $A_k$ be any planar 3-connected graph with $k$ faces, all of which are triangles,
and form a graph $G_k$ by replacing each triangle of $A_k$ by a copy of~$H$.  $G_k$ is thus a planar 3-connected graph with $n=37k$ faces, all of which are again triangles,
so it can be realized as a convex polyhedron with triangle faces.
The family described by the theorem consists of one such polyhedron for each $G_k$.

Now consider any single triangle strip formed by subdividing triangles of $G_k$.
With the exception of at most two copies of $H$ (the copies containing the start and end of the strip) the remaining $k-2$ copies of~$H$ must either be entered and exited exactly once by the strip (and therefore contain a subdivided edge in the interior of the copy, since $H$ has no cyclic triangle strip) or be entered and exited more than once (and therefore contain two subdivided edges on the triangle that was replaced by a copy of~$H$).  Thus, the total number of subdivided edges in the strip for $G_n$ is at least
$k-2$.  Each subdivided edge increases the number of triangles in the strip by two, so the total number of triangles in the strip is at least $39k-4=39n/37-4$.
\end{proof}

\subsection*{Lower Bounds for Manifolds with Boundaries}
Although our main results concern watertight models, we consider briefly for completeness the case of models with incomplete boundaries; we refer to any break in the boundary of a model as a {\em hole}.

% figure for section on lower bounds
\begin{figure*}[t]
\centering
\includegraphics[width=0.5\columnwidth]{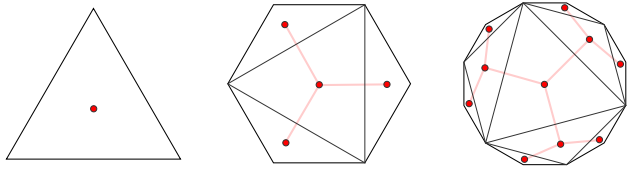}
\caption{Recursive triangulations of regular $(3\cdot2^k)$-gons, and their dual complete trees.}
\label{fig:holelb}
\end{figure*}

\begin{theorem}
\label{thm:holelb}
There exists an infinite family of triangulated models, having the topology of a sphere with a single hole, such that any single triangle strip formed by subdividing an $n$-triangle model in the family must
have at least $3n - 4\log_2 n-O(1)$ triangles.
\end{theorem}

\begin{proof}
Form model $M_k$ by triangulating a regular $(3\cdot 2^k)$-gon, so that there are $3\cdot 2^{k-1}$ outer triangles surrounding a central regular $(3\cdot 2^k)$-gon, which is triangulated recursively to form a copy of $M_{k-1}$.  As a base case for the recursion, let $M_0$ be a single equilateral triangle.
Figure~\ref{fig:holelb} depicts the first three models $M_0$, $M_1$, and $M_2$ in this sequence.
As shown in the figure, the dual graph of $M_k$ is a tree with $n_k=3(2^k-1)+1$ nodes, corresponding
to the same number of triangles in the model $M_k$.  The longest path in this tree has length~$2k$.

Now, consider any single triangle strip formed by subdividing $M_k$.  Any interior edge of $M_k$ that is not along the path from the start triangle to the end triangle of the strip must be crossed an even number of times by the strip, so all but $2k$ of the edges are subdivided and the total number of subdivided edges is at least $n_k - 1 - 2k$.
As in the previous theorem, each subdivision increases the number of triangles by two,
so the total number of triangles in the strip must be at least $3n_k - 2 - 4k=3n_k - 4\log_2 n_k - O(1)$.
\end{proof}

Finally, we show that the bound of Theorem~\ref{thm:holelb} is tight.

\begin{theorem}
Any connected $n$-triangle model with holes can be subdivided to form a triangle strip with
$3n-4\log_2 n+O(1)$ triangles in the strip.
\end{theorem}

\begin{proof} (sketch)
Let $T$ be any spanning tree of the dual graph of the model, and let $e$ be an edge in $T$ such that the two subtrees on each side of $e$ each have at least $n/3$ triangles in them; such an edge can be found by stepping from edge to edge towards the largest subtree until the condition is met.  Then,
the path $P$ formed by connecting the leaves farthest from $e$ in each subtree
has at least $\log_2 n/3 + \log_2 2n/3=2\log_2 n-O(1)$ edges.
Form a multigraph $M$ by doubling each edge of $T$ except for the edges in~$P$.
$M$~has even degree except at the endpoints of $P$, so it has an Euler path
that starts and ends at the endpoints of $P$.  If we add a new vertex at the midpoint of each internal edge of $T$ except for the edges in $P$, and subdivide each triangle of the model appropriately,
this Euler tour can be transformed into a triangle strip.  There are $n-2\log_2 n+O(1)$ new vertices,
so the total number of triangles in the strip is $3n-4\log_2 n + O(1)$.
\end{proof}

\end{document}